\documentclass[fleqn,10pt]{wlscirep}
\usepackage[utf8]{inputenc}
\usepackage[T1]{fontenc}
\usepackage{booktabs}
\title{Misinformation by Omission: \\ The Need for More Environmental Transparency in AI}
\author[1,*]{Sasha Luccioni}
\author[2]{Boris Gamazaychikov}
\author[3]{Theo Alves da Costa}
\author[4]{Emma Strubell}
\affil[1]{Hugging Face, Montreal, Canada}
\affil[2]{Salesforce, Paris, France}
\affil[3]{Ekimetrics, Paris, France}
\affil[4]{Carnegie Mellon University, School of Computer Science, Pittsburgh, USA}
\affil[*]{sasha.luccioni@huggingface.co}

\begin{abstract}
In recent years, Artificial Intelligence (AI) models have grown in size and complexity, driving greater demand for computational power and natural resources. In parallel to this trend, transparency around the costs and impacts of these models has decreased, meaning that the users of these technologies have little to no information about their resource demands and subsequent impacts on the environment. Despite this dearth of adequate data, escalating demand for figures quantifying AI's environmental impacts has led to numerous instances of misinformation evolving from inaccurate or de-contextualized best-effort estimates of greenhouse gas emissions. In this article, we explore pervasive myths and misconceptions shaping public understanding of AI's environmental impacts, tracing their origins and their spread in both the media and scientific publications. We discuss the importance of data transparency in clarifying misconceptions and mitigating these harms, and conclude with a set of recommendations for how AI developers and policymakers can leverage this information to mitigate negative impacts in the future.
\end{abstract}
\begin{document}

\flushbottom
\maketitle

\section*{Introduction}

AI-powered tools and systems are becoming increasingly ubiquitous, reshaping human behavior with corresponding impacts to our socioeconomic systems. While the companies that develop and deploy AI-driven technology strongly emphasize AI's positive impacts (real and speculative)~\cite{luers2024will,kshirsagar2021becoming,deepmind2023}, the negative impacts are typically left unspoken. These are often uncovered by researchers or a concerned public who audit AI systems, driven by the desire to understand the impacts of these systems on society and the planet.
Among the many impacts that have been analyzed in recent years are: algorithmic discrimination and bias~\cite{angwin2022machine, buolamwini2018gender}, the use of AI in military applications~\cite{rivera2024escalation}, the threat of AI to democracy~\cite{manheim2019artificial,summerfield2024will}, and environmental impacts~\cite{crawford2021atlas,bender2021dangers}.

Honing in on the latter, researchers and activists alike have been sounding the alarm on the increasingly unsustainable trends in energy and natural resource consumption arising from the data centers and devices used to train and deploy AI models which are growing in size and complexity~\cite{luccioni2022estimating, strubell2019energy,luccioni2023counting, dodge2022measuring,ligozat2021unraveling}. These resources range from the rare earth minerals necessary to manufacture computing hardware, the energy needed to power the very tangible ``cloud'' computation that underpins AI systems, the water needed for cooling and hardware manufacturing, and the greenhouse gases (GHGs) emitted at every stage of this process. Despite these rising costs, there exists minimal and often no data quantifying these impacts. When data does exist, it often lacks sufficient accessibility, detail and scope to enable effective decision-making or analysis, impeding impact assessment, mitigation, forecasting, and even basic understanding by the public at large. 

As a result, researchers, investors, companies and policymakers are left to attempt best-effort approximations given limited data availability.
In some cases, these estimates can be wildly flawed due to lack of critical prerequisite data, understanding or expertise. In other cases, estimates may be taken out of the careful, qualified contexts in which they were originally presented, leading to misinterpretation and in some cases severely inaccurate generalizations. The increasingly high stakes political and economic contexts surrounding climate change and AI severely compound this challenge; mistakes and misinterpretation devolve into misinformation as estimates are repeatedly shared and transformed through subsequent analyses, adopted as accurate measures and spread as trending posts through social media and the news, finally arriving on the desks of decision-makers. The resulting misconceptions harm all stakeholders: policymakers and the public are unable to make informed decisions, and AI technology developers suffer from negative perceptions arising from overestimates of their social harms, further exacerbating their lack of disclosure. In this paper, we aim to elucidate some common myths surrounding AI's environmental impacts, explore the pitfalls that led to the emergence of those myths, and propose recommendations for remedying this challenge in the future. 

\section*{Related Work} \label{section:related_work}

Approaches to calculating the environmental impacts of AI systems have evolved significantly over the last several years, as these systems have grown more impactful and widely deployed in user-facing applications. Initial studies, such as that of Strubell et al., underscored the environmental cost of training Transformer-based language models~\cite{strubell2019energy}. Research done in the following years extended this analysis, for instance by calculating the energy use and GHG footprint for several notable AI models including GPT-3, T5, Meena, and Switch Transformer, providing new estimates~\cite{patterson2021carbon} and expanding the scope of analysis beyond model training to account for operational and embodied emissions~\cite{gupta2021chasing}, improving methodology for software energy measurement \cite{cao-etal-2021-irene}, and a lifecycle approach to assessing emissions from model training and deployment~\cite{luccioni2022estimating}. Wu et al. further advanced this analysis by explicitly mapping the environmental impacts across the entire AI development pipeline,~\cite{wu2021sustainable}. Most recently, Luccioni, Jernite, and Strubell ~\cite{luccioni_2024} pioneered the AI inference impact methodology, revealing generative architectures as particularly energy-intensive compared to task-specific models and underscoring the critical importance of addressing inference impacts. These methodologies were then adapted into the AI Energy Score ~\cite{AIEnergyScore}, a project aiming to establish a unified approach for comparing the inference efficiency of AI models~\cite{luccioni2024light}. 

Above and beyond energy considerations, Li et al.~\cite{li2023making} expanded the scope of AI environmental impact measurement by estimating the water footprint of GPT-3 based on publicly available information, whereas Han et al.~\cite{han2024unpaid} assessed the public health toll of AI training’s air pollution, finding that training an AI model of the LLaMa 3.1 scale can produce air pollutants equivalent to more than 10,000 round trips by car between Los Angeles and New York City. In another significant advancement, Google's recent TPU lifecycle assessment~\cite{schneider2025lifecycleemissionsaihardware} offered the most comprehensive cradle-to-grave environmental analysis of AI hardware to date, integrating embodied carbon data associated with manufacturing AI accelerators and data center infrastructure, significantly extending existing environmental impact models. Building on many of these approaches, Morrison et al.~\cite{morrison2025holistically} performed a holistic evaluation of the energy, carbon, and water impacts of AI hardware manufacturing, model development, and training, enhancing the accuracy of these metrics through the use of granular underlying data.

The breadth and diversity of the analyses described in this section illustrate the multitude of factors involved in estimating AI’s environmental impacts, and the many different perspectives that exist in this space. Whereas several standardized approaches have been proposed to measure different aspects of AI’s requirements in terms of energy and water, as well as the emissions associated with model training and inference, the field is still currently lacking a comprehensive methodology and standards that cover all dimensions. In the next section, we examine how this translates into decreased environmental transparency in the AI industry via an empirical analysis of AI models over time. 

\section*{Environmental Transparency Trends} \label{section:trends}

\begin{figure}[h!]
  \centering    \includegraphics[width=\linewidth]{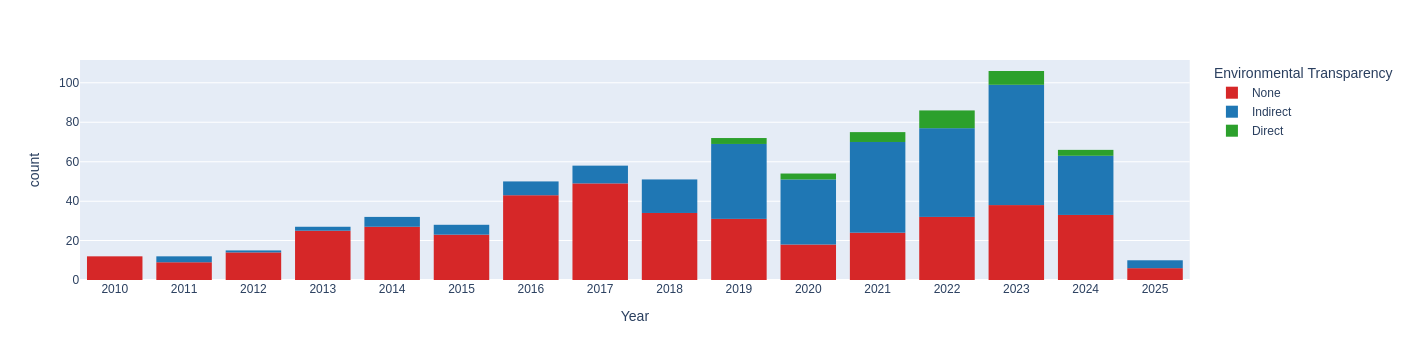}
  \caption{Environmental Impact Transparency of Notable AI Models by Release Year~\cite{EpochNotableModels2024}}
  \label{fig:transparency1}
\end{figure}

While there has been progress in developing more robust methodologies for measuring AI's environmental impacts, the broader AI industry has paradoxically been trending in the opposite direction, disclosing less information over time. In order to quantify this trend, we analyze Epoch AI's Notable AI Models dataset~\cite{EpochNotableModels2024}, which tracks information on \emph{``models that were state of the art, highly cited, or otherwise historically notable''}, with respect to transparency about the environmental impacts of those models. We examine the level of environmental impact transparency for each model based on key information from the Epoch AI dataset (e.g., model accessibility, training compute estimation method) as well as from individual model release content (e.g., paper, model card, announcement). We select the time period starting in 2010 as this is the beginning of the modern ``deep learning era'' (as defined by Epoch AI), which is representative of the types of AI models currently trained and deployed, including all 754 models from 2010 to the first quarter of 2025. Our analysis, shown in Figure~\ref{fig:transparency1}, reveals substantial variation in environmental impact transparency: some models disclose sufficient details to enable impact estimation, whereas others provide no information at all regarding their approach.

Overall, we find that models exhibit three transparency categories:
\begin{itemize}
\item \emph{Direct Disclosure}: Developers explicitly reported energy or GHG emissions.
Note that this category includes methodologies ranging from estimation (e.g., using hardware TDP, country average carbon intensity) to measurements (i.e., using tools like CodeCarbon). 
\item \emph{Indirect Disclosure}: Developers provided training compute data or released their model weights, allowing external estimates of training or inference impacts.
\item \emph{No Disclosure}: Environmental impact data was not publicly released and estimation approaches (as noted in Indirect Disclosure) were not possible.
\end{itemize}

From 2010 to 2018, only 17\% of the models shared data that could be used to indirectly estimate environmental impact of model training (ranging from 0 to 33\% each year); no direct environmental impact data was released during this period. This is expected, given that AI models of that era required significantly less compute and resource usage transparency was not yet common practice, although many articles accompanying papers did provide related information about, e.g. the amount of training data used or number of epochs trained. From 2019 to 2022, transparency improved as awareness of impacts grew and open-weights model releases became more common. This period includes the the work of Strubell et al.~\cite{strubell2019energy}, Luccioni~\cite{luccioni2022estimating} and others. The direct release of environmental information peaked in 2022, with 10\% of notable models that year releasing some degree of information. However, the introduction of increasingly commercial and proprietary models after 2022, potentially catalyzed by the popular launch of ChatGPT, which provided very limited information about the training approach used and even the final size of the underlying model, triggered a notable reversal in this trend, dramatically reducing direct environmental disclosures. By the first quarter of 2025, the majority of notable AI models again fell under the ``no disclosure'' category, as the line between research and commercial deployment became increasingly blurred.

\begin{figure}[h!]
  \centering
  \includegraphics[width=0.5\linewidth]{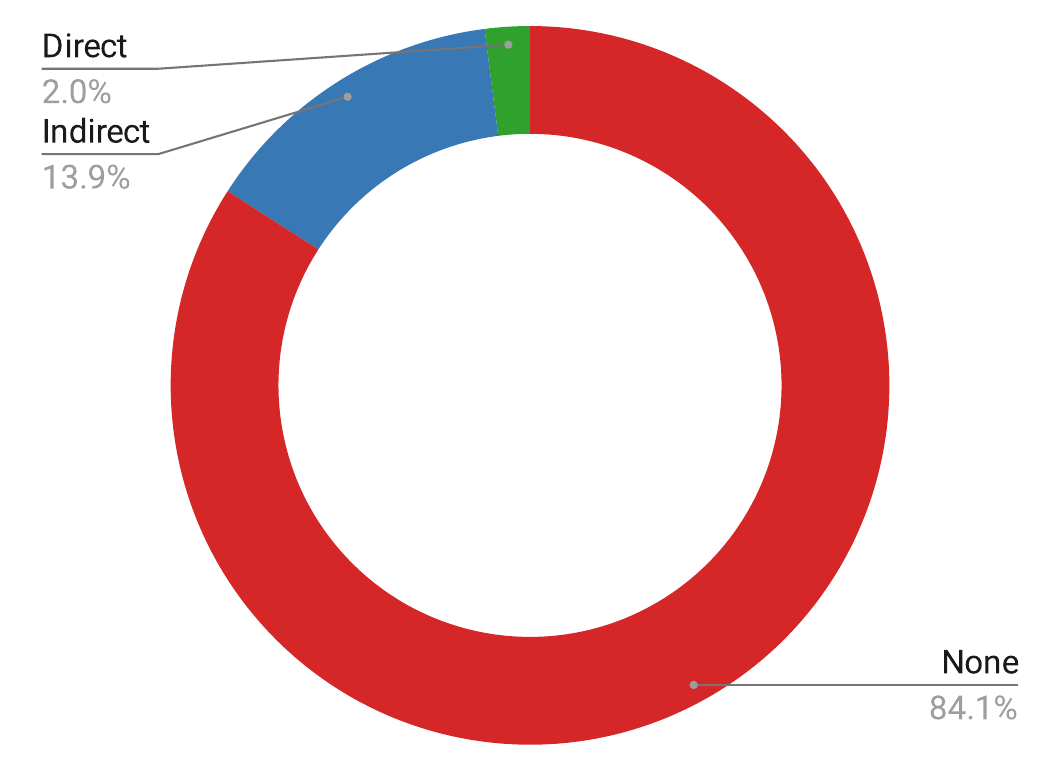}
  \caption{Environmental Impact Transparency of LLM Usage – OpenRouter~\cite{OpenRouterLeaderboard2025} (May 2025)}
  \label{fig:transparency2}
\end{figure}
Beyond the long term trend, zooming in to examine recent AI model usage data helps illustrate today’s environmental impact transparency conditions. OpenRouter~\cite{OpenRouterLeaderboard2025}, a widely-used API platform for LLMs, publicly shares data on model traffic including top 20 models by month, and the number of tokens running through every model. May 2025 data  (Figure~\ref{fig:transparency2}) indicates that of the top 20 used models, only one (Meta Llama 3.3 70B) directly released environmental data and three (DeepSeek R1, DeepSeek V3, Mistral Nemo) release it indirectly (by sharing compute data like GPU type and training length, as well as by releasing their model weights to enable efficiency analysis). In terms of token usage, 84\% of LLM usage is through models with no disclosure, 14\% for indirectly disclosed models, and only 2\% for models with direct disclosure. This indicates that the majority of users who interact with LLMs have no information about their environmental impacts, and cannot make informed decisions based on model efficiency or carbon intensity.

From the limited data that is publicly available, we can observe significant disparities in energy use and emissions across models. In fact, the energy required to pre-train an LLM spans from as little as 0.8 MWh (OLMo 20M) to 3,500 MWh (LLaMa 4 Scout), with associated GHG emissions varying even more significantly (due to variation in the carbon intensity of electricity across training locations). Inference workloads also show wide variation depending on model size, architecture and task type, with GPU energy usage for 1,000 queries spanning from just 0.06 Wh (bert-tiny) to over 3,426 Wh (Command-R Plus), depending on model size, architecture, and task complexity (see Tables~\ref{tab:pretraining} and \ref{tab:inference} in the Appendix for more information). These ranges highlight not only the scale of potential impacts, but also the pressing need for more standardized and transparent reporting to enable meaningful comparisons.

\section*{Investigating the Urban Legends of AI’s Environmental Impacts} \label{section:legends}

Making sustainably-minded decisions when using AI systems requires having the necessary information about different aspects of their development and deployment. While there are empirical studies focusing on AI’s environmental impacts, such as those cited in previous sections, these numbers have often been taken out of context or used as proxies for conditions (e.g., model size, architecture, optimizations, hardware, location, setup, system) that they are not representative of. This fuels misinformation, undermines scientific research, and can result in decisions that are not grounded in facts~\cite{lovins2025}. In the paragraphs below, we address some of the common estimates for the environmental impacts of AI, in an effort to contextualize their provenance and to explore their potential for spreading environmental misinformation.

\subsection*{Training an AI model emits as much CO$_2$ as five cars in their lifetimes}
Among the first efforts to quantify the environmental impacts of AI was the 2019 study by Strubell et al.,~\cite{strubell2019energy} which estimated the monetary costs, energy use, and GHG emissions required to train a variety of typical natural language processing (NLP) models of that era, including the first generation of large language models. This analysis included both the costs to train several individual models, including the two original ``base'' (65M) and ``big'' (213M parameter) variants of the Transformer neural network architecture~\cite{vaswani2017attention} that forms the basis of LLMs to this day, as well as the cost to perform model \textit{development}, i.e. identifying the best model architecture with respect to some optimization objective. The authors quantified the costs of model development through both a case study of the energy required for them to develop a model published in the previous year, and by estimating the energy required to automate that process using an approach called neural architecture search (NAS) based on figures reported in a recent Google study using NAS to identify an optimized variant of the Transformer architecture.\cite{pmlr-v97-so19a} In the case of the latter, they estimated that the NAS approach, assuming United States average electricity GHG emissions intensity and typical AI hardware running in an average-efficiency datacenter, could yield 626,155 pounds (284 metric tons) CO$_2$-equivalent GHG emissions (CO$_2$e), or about five times the emissions of a car during its lifetime, including fuel.

The research article was written for a specialized audience of AI and NLP researchers, who would have the background knowledge to understand the appropriate scoping for the estimate. However, an author's tweet publicizing the paper and featuring a table containing the ``five cars'' estimate was widely shared on social media, leading to the publication being picked up by numerous media outlets (including MIT Technology Review~\cite{hao2019training} and Forbes~\cite{toews2020}). The ``five cars'' number has since been misinterpreted as a proxy for the carbon footprint of training AI models at large, which is misleading given the diversity of architectures, training approaches and electricity sources used for powering AI model training; the original article reports AI training workloads emitting as little as 26 pounds (11.8 kg) CO$_2$e (assuming U.S. average energy carbon emissions intensity), and AI model training more broadly often requires even less energy and corresponding emissions. 

Further, the NAS training workload represents a large-scale procedure that is meant to be and is in practice performed much less frequently than the average AI model training workload. This is both because the result is intended to be re-used as a basis to reduce the emissions of subsequent training workloads, and because the scale of resources (financial and/or computational) significantly limits who can perform such large-scale training runs. In this way, the NAS training workload is similar to today's generative AI pretraining workloads, which are similarly performed less frequently than the average AI training. 
However, while the ``five cars'' estimate from Strubell et al. is not an accurate representation of the emissions arising from every AI training workload, recent first-hand reports of the estimated GHG emissions arising from language model pretraining typically exceed the ``five cars'' estimate: Google reports that training their open source Gemma family of language models emitted 1247.61 tons CO$_2$e,\cite{gemmateam2024gemma2improvingopen} over 4x the estimate that forms the basis for the ``five cars'' number, and Meta reports that their Llama 3 family of models emitted 11,390 tons CO$_2$e~\cite{Llama31ModelCard} or over 40x the ``five cars'' estimate.

\subsection*{A request to ChatGPT consumes ten times more energy than a Google search}

Another often cited and misrepresented metric is the estimate that a single request to ChatGPT uses approximately 3 watt-hours (Wh) of energy, which is "ten times more than a Google search". This figure is often quoted in the press~\cite{kerr2024ai, chen2025much} and in industry reports~\cite{aljbour2024powering}. Tracing the origins of this metric leads to several assumptions: an initial remark from Alphabet's Chairman John Hennessy during a 2023 interview with Reuters, in which he said that \emph{``having an exchange with AI known as a large language model likely cost 10 times more than a standard keyword search''}~\cite{dastin2023focus}. This remark was used was the basis of an estimate published in October 2023 of \emph{``approximately 3 Wh per LLM interaction''}~\cite{de2023growing}, with the Google search number taken from a 2009 blog post from Google that stated that \emph{``Queries vary in degree of difficulty, but for the average query [...] this amounts to 0.0003 kWh of energy per search''~\cite{google2009}}. This number is misleading for several reasons. First, Hennessy has no relation to OpenAI or Microsoft (which provides the compute for OpenAI's services), so the comment he made was based on secondhand information. Second, even if Hennessy's comparison were accurate, basing the search estimate on a figure that is 16 years old --- at a time when Web search was done using bag-of-words or vector-based search techniques as opposed to the current Transformer-based models --- is also bound to amplify the inaccuracy of the estimate.

\begin{figure}[ht]
\centering
\includegraphics[width=0.8\linewidth]{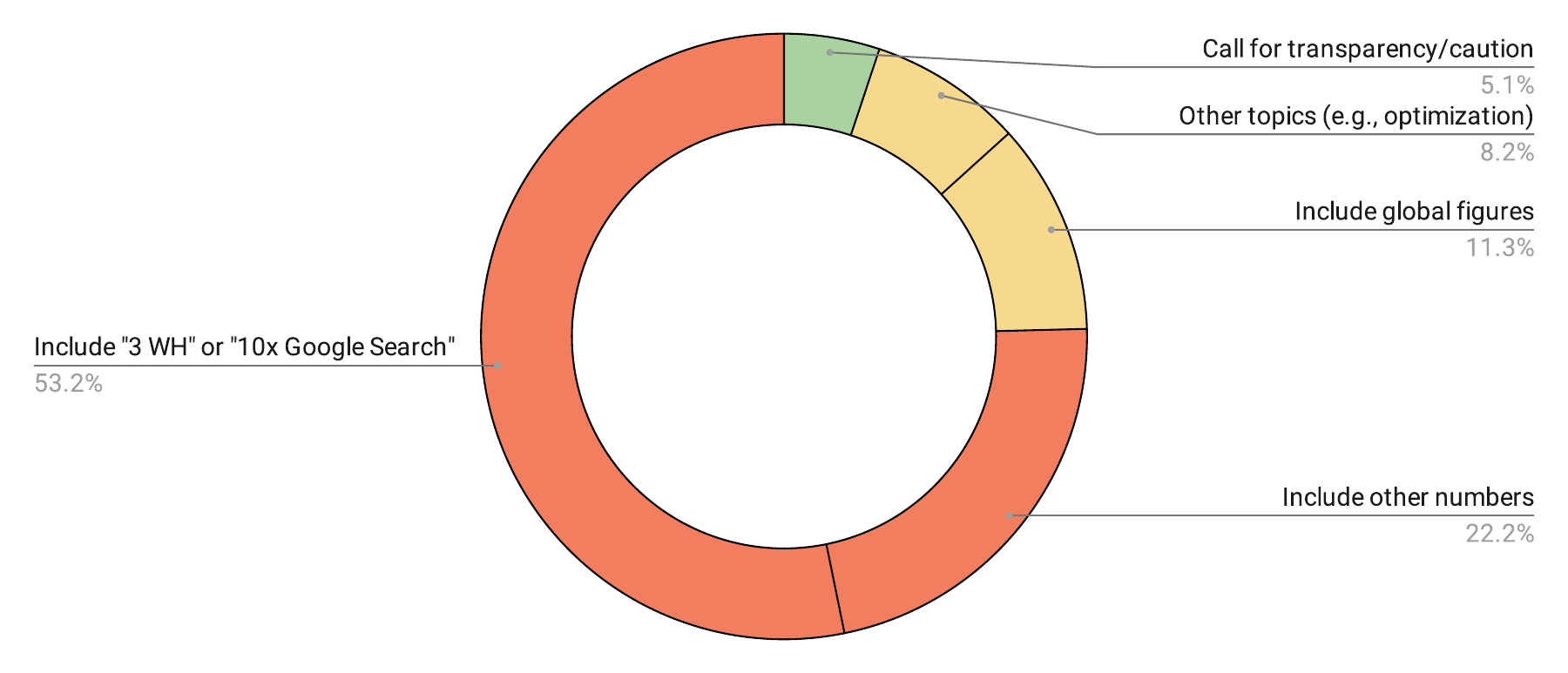}
\caption{Analysis of media articles discussing ChatGPT energy consumption.}
\label{fig:media}
\end{figure}

To understand the impact of the propagation of this estimate, we analyzed 100 news articles published as of April 11, 2025, that appear when searching for \emph{``ChatGPT energy consumption''} on Google News. For each article, we noted whether it mentioned the 3 Wh estimate, if it referenced others, or if it called for transparency or caution regarding figures by acknowledging uncertainty or suggesting that such statistics should be viewed critically. Our results, shown in Figure~\ref{fig:media}, reveal that 75\% of media articles relayed energy estimates for a ChatGPT query without mentioning uncertainties or even citing the sources for these figures: 53\% of articles cite the figure of 3 Wh per ChatGPT query or claim it consumes 10 times more energy than a Google search\cite{euronews}, 22\% mention other precise energy numbers for ChatGPT queries, comparing them to the number of American households or LED light bulbs~\cite{theconversation} (likely using the same 3 Wh figure), 11\% prefer to provide global figures on the energy impact of data centers\cite{cnbc}, 8\% discuss other topics, particularly DeepSeek~\cite{mittechreview} and optimizations with ternary neural network architectures to improve energy efficiency~\cite{techzine} and only 5\% explicitly call for transparency or necessary caution when addressing this subject~\cite{vox}, stating that the true figures remain unknown.
It is also noteworthy that among these articles,
9\% also relay the claim that training a LLM produces emissions equivalent to 5 cars in their lifetime.

\subsection*{AI can reduce 10\% of global emissions}

While the numbers around AI's negative environmental impacts can be misinterpreted and taken out of context, so, too, can the potential of AI to reduce emissions, especially by corporate actors that develop and deploy AI systems on a global scale. One recurring number states that AI can help reduce global GHG emissions (up to) 10\%. This number can be traced back to a 2021 Boston Consulting Group (BCG) report which states that \emph{``Research shows that by scaling currently proven applications and technology, AI could mitigate 5 to 10\% of global greenhouse gas emissions by 2030–the equivalent of the total annual emissions of the European Union''}~\cite{degot2021reduce}. The same number appears in a more recent BCG report from 2023, which was commissioned by Google and published ahead of COP26~\cite{dannouni2023accelerating}. The reasoning behind the 5-10\% reduction estimate is unclear and the underlying calculations are not detailed beyond the explanation that they are based on BCG's experience in dealing with their clients and using AI to optimize and improve existing processes. 
The second, Google-commissioned BCG study provides slightly more detail in terms of the kinds of projects AI can be used for, but does not offer specific calculations translating individual project numbers to a global scale. 

Applying observations made from individual projects to the entire planet's GHG emissions lacks any scientific grounding---in fact, many of the emissions reductions on a global scale require individual, societal and political shifts. 
Moreover, rigorous calculation of avoided emissions requires defining counterfactual reference scenarios, conducting systematic consequence analysis, and accounting for rebound effects---methodological requirements outlined in established recent standards like ITU-T L.1480\cite{itu-methodo} or WBCSD guidance on avoided emissions\cite{wbcsd-methodo}.  And yet, these numbers were picked up in research~\cite{das2023survey} and the media, used as evidence that the potential of AI to stop climate change is overwhelmingly positive~\cite{theenvironment2022,kakkad2023}. 
While AI undoubtedly has potential positive applications in sectors ranging from transportation to agriculture to energy~\cite{rolnick2022tackling}, these global generalizations can be misleading because they overlook the myriad of problems that technology alone cannot solve, while giving credibility to the beliefs that the benefits of AI will outweigh its costs~\cite{guardian2024}.

The lack of transparency around AI’s environmental impacts can have far-reaching consequences, ranging from specific estimates taken out of context and blown out of proportion, to proxies becoming adopted by press and policymakers in the absence of more reliable figures. In the next section, we discuss a potential solution to this situation by proposing a set of metrics that different stakeholders can measure and report to bring more clarity to the extent of AI’s environmental impacts. 

\section*{How to improve environmental impact disclosures in AI}
Opacity in AI environmental reporting creates multiple interconnected challenges: organizations cannot make informed procurement or innovation decisions without access to reliable environmental performance data on AI, while policymakers lack the information necessary to develop evidence-based regulations. This opacity also generates cascading effects throughout value chains, as AI adoption creates unmeasured emissions that undermine corporate net zero commitments. Furthermore, the absence of standardized metrics prevents meaningful comparison between AI systems, limiting market mechanisms that could drive efficiency improvements. Perhaps most critically, this lack of transparency undermines accountability mechanisms, making it impossible to hold AI developers and deployers responsible for their environmental performance or to track progress toward sustainability goals.

This section explores how comprehensive environmental transparency can address these challenges through four interconnected pathways:
\begin{enumerate}
\item Carrying out comprehensive measurement and disclosure by AI developers at each stage of model development and deployment;
\item Integrating comprehensive AI environmental impacts into sustainability accounting frameworks and corporate sustainability disclosures by organizations across the entire AI value chain, from model providers and hyperscalers to end-user enterprises;
\item Developing standardized verification and assurance frameworks to ensure data reliability and enable meaningful comparisons; and
\item Implementing clear regulatory requirements by policymakers to ensure consistent, verifiable reporting across the industry.
\end{enumerate}

\paragraph{Measurement and Disclosure} As the starting point of AI development, AI researchers and developers are able to gather empirical measurements from the systems they create at different steps of the model lifecycle. When developing models from scratch, energy consumption and GHG emissions from training and inference can be estimated using programmatic tools like Code Carbon~\cite{schmidt2021codecarbon} or no-code tools like Green Algorithms~\cite{lannelongue2021green}. When using or adapting existing models, performance and efficiency testing can significantly reduce emissions by enabling the deployment of more energy-efficient models in production. For instance, the AI Energy Score project~\cite{AIEnergyScore} provides a standardized methodology for comparing models across different tasks, which can also be adapted for specific contexts and datasets. These metrics should be reported in model cards~\cite{mitchell2019model} and scientific publications with complete methodological transparency, including hardware specifications, geographic locations, electricity sources, measurement uncertainties, and allocation methodologies. This empirical foundation enables downstream organizational GHG accounting while contributing to the broader scientific understanding of AI environmental impacts through peer-reviewed publication of methodologies and results. AI providers across the entire value chain, including cloud infrastructure providers, model hosting platforms, and API service providers, must implement comprehensive transparency with granular environmental data disclosure, enabling downstream organizations to accurately account for their AI-related environmental impacts. Government and public sector organizations should mandate transparency in all AI procurements, require open data for publicly funded research, and align AI deployments with existing net zero commitments.

\paragraph{Organizational Implementation and Processes}
As AI adoption accelerates, organizations should implement comprehensive frameworks to assess, measure, and integrate AI’s environmental impacts into existing sustainability management systems using structured approaches tailored to their specific contexts and risk profiles. The materiality assessment framework should aim to establish quantitative thresholds across environmental intensity and usage scale dimensions -- for example creating distinct tiers of analysis intensity. 
Organizations developing AI systems utilizing open-source models on their infrastructure should implement comprehensive measurement protocols at multiple levels of granularity: model-specific, service or process-level, and organization-wide aggregations. 
Similarly, entities utilizing third-party AI services (e.g., API-based integrations of commercial models or subscription-based access for internal teams like ChatGPT, Copilot or Claude) should demand transparency by incorporating environmental disclosure requirements into procurement processes and contractual agreements~\cite{AIES_call_to_action2025}. 
Specifically, organizations should request access to standardized metrics (such as the AI Energy Score or an equivalent) for all AI services under consideration.
These environmental metrics should be systematically integrated into organizations' GHG accounting frameworks and non-financial performance disclosures, with explicit documentation of methodological assumptions and unmodeled factors. 

\paragraph{Standards, Verification and Assurance} Environmental AI disclosures require robust verification frameworks to ensure accuracy and prevent greenwashing, necessitating new assurance standards adapted to AI's rapid evolution, distributed compute, and complex value chains. While no unified standard yet exists for assessing AI sustainability, parallel efforts are underway across organizations such as the Green Software Foundation, ISO (International Organization for Standardization), and OECD (Organization for Economic Cooperation and Development). These bodies are well-positioned to develop standardized approaches for stakeholders ranging from developers to governments. Given AI's transnational nature, coordination and harmonization of these efforts is essential. Without alignment, implementation may diverge across jurisdictions, creating further confusion in the market. However, as formal standards may take years to materialize, interim ad hoc methods (such as those outlined above) can provide valuable insights and help shape the eventual development of formal methodologies. 
These AI environmental disclosure frameworks must also strengthen adherence to robust GHG accounting principles, particularly regarding the GHG Protocol's treatment of electricity emissions measurement. The current allowance for market-based accounting enables companies to significantly under-report their actual AI-related emissions through renewable energy certificates, creating the same problematic disconnect from reality that has undermined carbon offsetting credibility ~\cite{guardian2024}. For AI services consuming substantial electricity across distributed data centers, mandatory location-based accounting would ensure environmental transparency frameworks capture the true systemic climate impacts rather than allowing them to be obscured through market mechanisms.

\paragraph{Policy Frameworks and Reporting}
Environmental transparency documentation is already commonplace for private organizations in existing legislation such as the Corporate Sustainability Reporting Directive (CSRD) in the EU, SEC climate disclosure requirements in the US, or local and state-level climate disclosure laws. However, policymakers should incorporate additional reporting requirements specifically addressing AI system utilization under standards such as European Sustainability Reporting Standards E1 (Climate Change) which mandates the disclosure of Scope 1, 2, and 3 GHG emissions, energy usage, and a transition plan aligned with the Paris Agreement~\cite{leal2025european}, particularly as this aligns with existing provisions in the EU AI Act. 
Non-governmental sustainability rating agencies such as CDP and EcoVadis should similarly expand their assessment criteria to incorporate AI-specific environmental impact metrics, creating market incentives for improved disclosure practices.
For organizations directly participating in the AI value chain (service providers, data center operators, developers, IT integrators, semiconductor and GPU manufacturers) policymakers should implement more stringent transparency requirements. These could include mandatory detailed environmental reporting disaggregated by model, usage patterns, and physical infrastructure. Enforcement mechanisms might include annual comprehensive environmental reports or conditioning access to public markets and funding on compliance with disclosure standards. 

\section*{Conclusion}

The current trend toward reduced transparency around AI’s environmental impact contributes to misinformation and hinders informed decision-making across all levels, from individual researchers and developers to organizations and policymakers. This declining transparency is particularly troubling given AI's escalating environmental impacts amid global climate concerns and looming planetary boundaries. While competition is frequently cited to justify opacity, other competitive industries, such as food (with ingredient labeling) and healthcare (with side-effect and pricing transparency), demonstrate that a balance between transparency and competition is achievable. Reversing the trend toward opacity in AI environmental reporting is essential for informed decision-making, accountability, and sustainable technology advancement, particularly as new model paradigms emerge that may alter these impacts. As members of the AI community committed to addressing the climate crisis, we aim to ensure the sustainability of our field as it continues to expand -- recognizing that increased transparency is fundamental to this goal.

\clearpage
\bibliography{biblio}



\section*{Author contributions statement}
B.G. conducted the environmental impact transparency analysis, T.A.d.C carried out the media analysis. All authors wrote, edited and reviewed the manuscript.


\clearpage
\section*{Appendix}

\begin{table}[h!]

\centering
\caption{Range of Pre-Training Environmental Impacts (Representative Models Displayed)}
\label{tab:pretraining}
\resizebox{\textwidth}{!}{%
\begin{tabular}{cccc}
\hline
\multicolumn{1}{|c|}{\textbf{Model}} & \multicolumn{1}{c|}{\textbf{Organization}} & \multicolumn{1}{c|}{\textbf{Energy Consumption (MWh)}} & \multicolumn{1}{c|}{\textbf{GHG Emissions (tCO2e)}} \\ \hline
\multicolumn{1}{|c|}{OLMo 20M~\cite{morrison2025holistically}}       & \multicolumn{1}{c|}{Ai2}                   & \multicolumn{1}{c|}{0.8}                               & \multicolumn{1}{c|}{0.3}                               \\ \hline
\multicolumn{1}{|c|}{CodeGen 350M~\cite{SalesforceBlueprint}}   & \multicolumn{1}{c|}{Salesforce}            & \multicolumn{1}{c|}{71}                                & \multicolumn{1}{c|}{6}                                 \\ \hline
\multicolumn{1}{|c|}{Llama 7B~\cite{touvron2023llamaopenefficientfoundation}}       & \multicolumn{1}{c|}{Meta}                  & \multicolumn{1}{c|}{356}                               & \multicolumn{1}{c|}{14}                                \\ \hline
\multicolumn{1}{|c|}{BLOOM~\cite{luccioni2022estimating}}          & \multicolumn{1}{c|}{Big Science}           & \multicolumn{1}{c|}{520}                               & \multicolumn{1}{c|}{30}                                \\ \hline
\multicolumn{1}{|c|}{T5~\cite{patterson2021carbon}}             & \multicolumn{1}{c|}{Google}                & \multicolumn{1}{c|}{85.7}                              & \multicolumn{1}{c|}{47}                                \\ \hline
\multicolumn{1}{|c|}{OLMo 2 13B~\cite{morrison2025holistically}}     & \multicolumn{1}{c|}{Ai2}                   & \multicolumn{1}{c|}{157}                               & \multicolumn{1}{c|}{101}                               \\ \hline
\multicolumn{1}{|c|}{Gemma 2B + 9B~\cite{gemmateam2024gemmaopenmodelsbased}}  & \multicolumn{1}{c|}{Google}                & \multicolumn{1}{c|}{?}                                 & \multicolumn{1}{c|}{131}                               \\ \hline
\multicolumn{1}{|c|}{GPT-3~\cite{patterson2021carbon}}          & \multicolumn{1}{c|}{OpenAI}                & \multicolumn{1}{c|}{1,287}                             & \multicolumn{1}{c|}{552}                               \\ \hline
\multicolumn{1}{|c|}{Llama 4 Scout~\cite{Llama4ModelCard}}  & \multicolumn{1}{c|}{Meta}                  & \multicolumn{1}{c|}{3,500}                             & \multicolumn{1}{c|}{1,354}                             \\ \hline
\multicolumn{1}{|c|}{Llama 3 70B~\cite{Llama3ModelCard}}    & \multicolumn{1}{c|}{Meta}                  & \multicolumn{1}{c|}{?}                                 & \multicolumn{1}{c|}{1,900}                             \\ \hline
\multicolumn{1}{|c|}{Llama 3.1 405B~\cite{Llama31ModelCard}} & \multicolumn{1}{c|}{Meta}                  & \multicolumn{1}{c|}{?}                                 & \multicolumn{1}{c|}{8,930}                             \\ \hline
\multicolumn{1}{l}{}                 & \multicolumn{1}{l}{Max/Min Variance:}      & 4,375                                                  & 29,767                                                
                                     
\end{tabular}
}
\end{table}

\begin{table}[]
\centering
\caption{Range of Inference Energy Use\cite{AIEnergyScore} (Representative Models Displayed)}
\label{tab:inference}
\resizebox{\textwidth}{!}{%
\begin{tabular}{cccc}
\hline
\multicolumn{1}{|c|}{\textbf{Model}}              & \multicolumn{1}{c|}{\textbf{Organization}} & \multicolumn{1}{c|}{\textbf{GPU Energy for 1k Queries (Wh)}} & \multicolumn{1}{c|}{\textbf{Task}}        \\ \hline
\multicolumn{1}{|c|}{bert-tiny-finetuned-squadv2} & \multicolumn{1}{c|}{mrm8488}               & \multicolumn{1}{c|}{0.06}                                    & \multicolumn{1}{c|}{Extractive QA}        \\ \hline
\multicolumn{1}{|c|}{GIST-all-MiniLM-L6-v2}       & \multicolumn{1}{c|}{avsolatorio}           & \multicolumn{1}{c|}{0.11}                                    & \multicolumn{1}{c|}{Sentence Similarity}  \\ \hline
\multicolumn{1}{|c|}{dynamic\_tinybert}           & \multicolumn{1}{c|}{Intel}                 & \multicolumn{1}{c|}{0.21}                                    & \multicolumn{1}{c|}{Extractive QA}        \\ \hline
\multicolumn{1}{|c|}{distilbert-imdb}             & \multicolumn{1}{c|}{lvwerra}               & \multicolumn{1}{c|}{0.22}                                    & \multicolumn{1}{c|}{Text Classification}  \\ \hline
\multicolumn{1}{|c|}{question\_answering\_v2}     & \multicolumn{1}{c|}{Falconsai}             & \multicolumn{1}{c|}{0.23}                                    & \multicolumn{1}{c|}{Extractive QA}        \\ \hline
\multicolumn{1}{|c|}{Resnet 18}                   & \multicolumn{1}{c|}{Microsoft}             & \multicolumn{1}{c|}{0.30}                                    & \multicolumn{1}{c|}{Image Classification} \\ \hline
\multicolumn{1}{|c|}{yolos-tiny}                  & \multicolumn{1}{c|}{hustvl}                & \multicolumn{1}{c|}{1.00}                                    & \multicolumn{1}{c|}{Object Detection}     \\ \hline
\multicolumn{1}{|c|}{Vision Perceiver Conv}       & \multicolumn{1}{c|}{Google}                & \multicolumn{1}{c|}{2.64}                                    & \multicolumn{1}{c|}{Image Classification} \\ \hline
\multicolumn{1}{|c|}{SFR-Embedding-Mistral}       & \multicolumn{1}{c|}{Salesforce}            & \multicolumn{1}{c|}{5.22}                                    & \multicolumn{1}{c|}{Sentence Similarity}  \\ \hline
\multicolumn{1}{|c|}{yolos-base}                  & \multicolumn{1}{c|}{hustvl}                & \multicolumn{1}{c|}{7.98}                                    & \multicolumn{1}{c|}{Object Detection}     \\ \hline
\multicolumn{1}{|c|}{Gemma 7B}                    & \multicolumn{1}{c|}{Google}                & \multicolumn{1}{c|}{18.90}                                   & \multicolumn{1}{c|}{Text Generation}      \\ \hline
\multicolumn{1}{|c|}{T5 11b}                      & \multicolumn{1}{c|}{Google}                & \multicolumn{1}{c|}{27.79}                                   & \multicolumn{1}{c|}{Text Classification}  \\ \hline
\multicolumn{1}{|c|}{phi-4}                       & \multicolumn{1}{c|}{Microsoft}             & \multicolumn{1}{c|}{28.74}                                   & \multicolumn{1}{c|}{Text Generation}      \\ \hline
\multicolumn{1}{|c|}{T5 11b}                      & \multicolumn{1}{c|}{Google}                & \multicolumn{1}{c|}{178.13}                                  & \multicolumn{1}{c|}{Extractive QA}        \\ \hline
\multicolumn{1}{|c|}{Mitsua Diffusion One}        & \multicolumn{1}{c|}{Mitsua}                & \multicolumn{1}{c|}{186.81}                                  & \multicolumn{1}{c|}{Image Generation}     \\ \hline
\multicolumn{1}{|c|}{Mixtral 8x7B}                & \multicolumn{1}{c|}{Mistral}               & \multicolumn{1}{c|}{615.39}                                  & \multicolumn{1}{c|}{Text Generation}      \\ \hline
\multicolumn{1}{|c|}{Stable Diffusion XL Base}    & \multicolumn{1}{c|}{Stability AI}          & \multicolumn{1}{c|}{1,639.85}                                & \multicolumn{1}{c|}{Image Generation}     \\ \hline
\multicolumn{1}{|c|}{Llama 3 70B}                 & \multicolumn{1}{c|}{Meta}                  & \multicolumn{1}{c|}{1,719.66}                                & \multicolumn{1}{c|}{Text Generation}      \\ \hline
\multicolumn{1}{|c|}{Qwen2.5 72B}                 & \multicolumn{1}{c|}{Qwen}                  & \multicolumn{1}{c|}{1,869.55}                                & \multicolumn{1}{c|}{Text Generation}      \\ \hline
\multicolumn{1}{|c|}{Command-R Plus}              & \multicolumn{1}{c|}{Cohere}                & \multicolumn{1}{c|}{3,426.38}                                & \multicolumn{1}{c|}{Text Generation}      \\ \hline
\multicolumn{1}{l}{}                              & \multicolumn{1}{l}{Max/Min Variance:}      & 57,106                                                       & \multicolumn{1}{l}{}                     
\end{tabular}
}
\end{table}

\end{document}